\documentclass[sigconf]{acmart}
\AtBeginDocument{%
  \providecommand\BibTeX{{%
    \normalfont B\kern-0.5em{\scshape i\kern-0.25em b}\kern-0.8em\TeX}}}

\copyrightyear{2022}
\acmYear{2022}
\setcopyright{acmcopyright}\acmConference[ICSE-NIER'22]{New Ideas and Emerging Results }{May 21--29, 2022}{Pittsburgh, PA, USA}
\acmBooktitle{New Ideas and Emerging Results (ICSE-NIER'22), May 21--29, 2022, Pittsburgh, PA, USA}
\acmPrice{15.00}
\acmDOI{10.1145/3510455.3512771}
\acmISBN{978-1-4503-9224-2/22/05}

\usepackage{framed}
\usepackage{enumitem}

\begin{document}

\title{Better Modeling the Programming World with Code Concept Graphs-augmented Multi-modal Learning}


\author{Martin Weyssow}
\affiliation{%
\institution{DIRO, Universit\'e de Montr\'eal}
\city{Montreal}
\country{Canada}}
\email{martin.weyssow@umontreal.ca}

\author{Houari Sahraoui}
\affiliation{%
\institution{DIRO, Universit\'e de Montr\'eal}
\city{Montreal}
\country{Canada}}
\email{sahraouh@iro.umontreal.ca}

\author{Bang Liu}
\affiliation{%
\institution{DIRO \& Mila, Universit\'e de Montr\'eal}
\city{Montreal}
\country{Canada}}
\email{bang.liu@umontreal.ca}



\begin{abstract}
The progress made in code modeling has been tremendous in recent years thanks to the design of natural language processing learning approaches based on state-of-the-art model architectures. Nevertheless, we believe that the current state-of-the-art does not focus enough on the full potential that data may bring to a learning process in software engineering. Our vision articulates on the idea of leveraging multi-modal learning approaches to modeling the programming world. In this paper, we investigate one of the underlying idea of our vision whose objective based on concept graphs of identifiers aims at leveraging high-level relationships between domain concepts manipulated through particular language constructs. In particular, we propose to enhance an existing pretrained language model of code by joint-learning it with a graph neural network based on our concept graphs. We conducted a preliminary evaluation that shows gain of effectiveness of the models for code search using a simple joint-learning method and prompts us to further investigate our research vision. 
\end{abstract}



\maketitle

\section{Introduction}
\label{sec:introduction}

One of the main objective of language models and graph neural networks (GNN) in natural language processing (NLP) is to implicitly encode knowledge about the world by learning semantically meaningful representations of words and concepts. In software engineering, language models have been extensively used to modeling the programming world~\cite{hindle2016naturalness, allamanis2013mining, feng2020codebert} and reason about it through a variety of tasks~\cite{allamanis2018survey, austin2021program}. Alternatively, graph neural networks have been used to learn from structured information such as abstract syntax trees and have shown great success to modeling source code~\cite{allamanis2017learning, brockschmidt2018generative, hellendoorn2019global}. 
Most of the aforementioned related work focus on learning either from a unique abstraction of a data source or from multiple data sources, \textit{e.g.,} a code and its corresponding documentation. We claim that neither of these approaches are sufficient to fully modeling the programming world and that we need to design multi-modal approaches tailored to learn from different levels of abstractions of the same software artifact, \textit{e.g.,} a source code and its AST. Furthermore, we conjecture that high-level data abstractions such as concept graphs (CG) could guide learning models towards a better understanding of software.
\begin{figure}[!t]
    \centering
    \includegraphics[width=\linewidth]{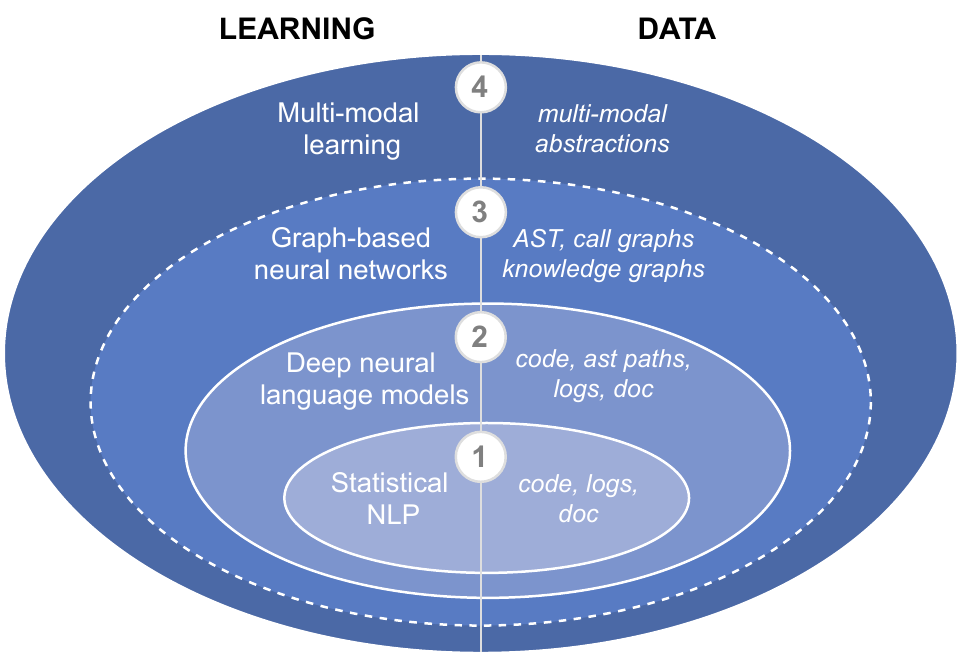}
    \caption{NLP-based learning approaches in software engineering research with their corresponding usual data sources. The outermost layer relates to our vision about combining multiple data modalities and learning models.} 
    \label{fig:abstraction}
\end{figure}

\begin{figure*}[!ht]
    \centering
    \includegraphics[width=.45\textwidth]{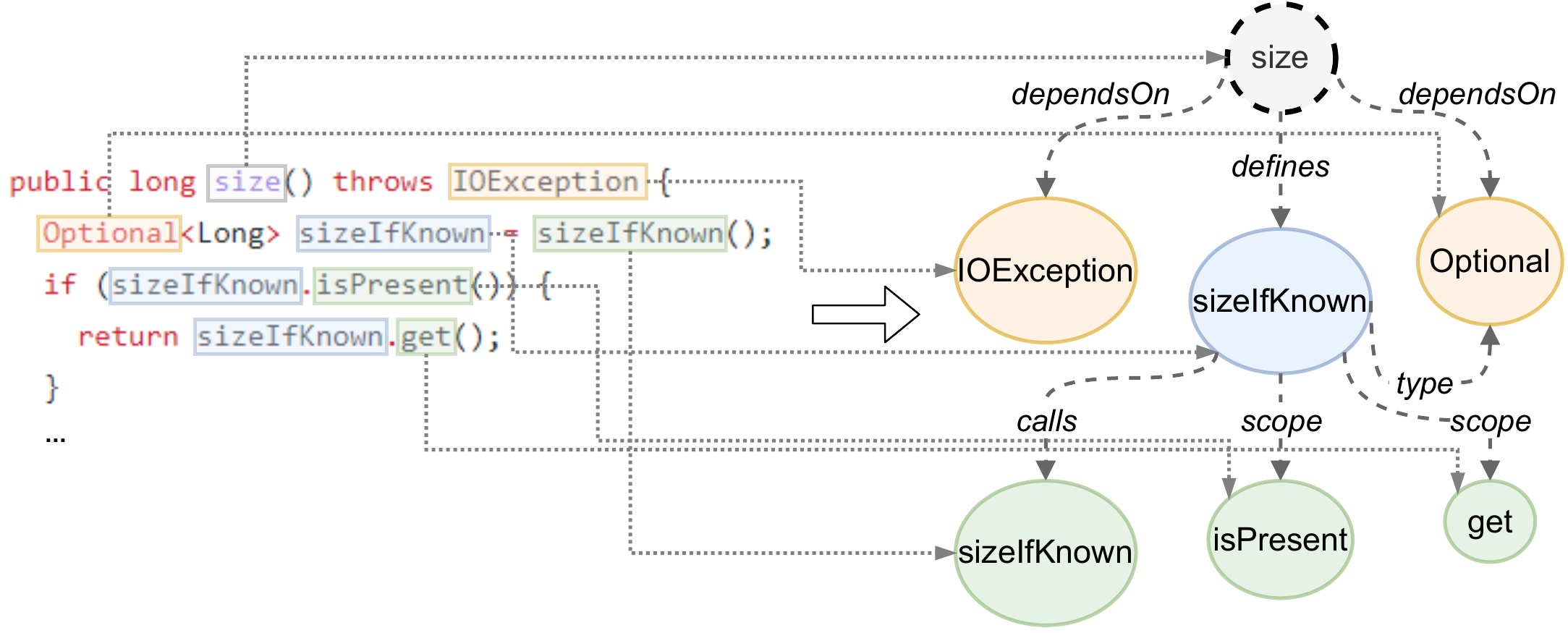}
    \caption{Excerpt of the extraction of a concept graph from a code snippet in Java.}
    \label{fig:concept_graph}
\end{figure*}

In fact, as we human communicate by the mean of natural languages, we manipulate words and concepts at different level of abstraction. That is, our understanding of the language relies on a principle of compositionality where high-level concepts are defined as compositions of lower level concepts, \textit{e.g.,} a tree is made of leaves and a trunk. Furthermore, at the grammar-level of the language, a sentence can be decomposed as a dependency graph between words or group of words making the language very hierarchical and structured~\cite{marcus1993building,jurafsky2000speech,tai2015improved}. In a similar way, this notion of compositionality and hierarchies is central in programming languages. Even though we write code as sequences of textual elements, these sequences have an equivalent deterministic abstraction in form of an abstract syntax tree that encodes structural, semantic and syntactic information of the code. In other terms, AST-like abstractions of the code have the benefit of incorporating higher level information such as the logical structure of code. In addition and as part of this work, analogously to natural languages we conjecture that programs are concept-based. That is to say, a program can be decomposed in term of high-level concerns, \textit{e.g.,} a method or a class, that manipulate low-level concepts which target a more specific and local purpose such as a variable. In addition, programming concepts are not always defined as such and can designate a direct manipulation of domain concepts, \textit{e.g.,} specified in the requirements of the software.
In other words, we underline the fact that concepts are deeply rooted in source code artifacts and manipulated using particular language-specific constructs each of them associated with an identifier bearing the meaning of the concept. 

In order to support our first claim, lets go through a discussion about the first three layers depicted in Fig.~\ref{fig:abstraction} which summarizes NLP-based learning approaches widely used to learn from text with concrete examples of data sources specific to software engineering. 
The first two layers relate to token-based language modeling, \textit{i.e.,} learning from sequences of texts. On the one hand, statistical NLP approaches are based upon learning recurrent patterns from software artifacts such as code, commit logs or natural language documentation~\cite{hindle2016naturalness, allamanis2013mining, hellendoorn2017deep}. On the other hand, recent deep-learning-based approaches leveraging powerful language model architectures~\cite{hochreiter1997long, bahdanau2014neural, vaswani2017attention, devlin2018bert, radford2018improving} to learn representations of source code have shown impressive results in tasks such as code completion or code search~\cite{raychev2014code, karampatsis2020big, svyatkovskiy2021fast, feng2020codebert, lu2021codexglue}. Within these two kind of approaches, the learning model relies heavily on linguistics features specific to the input language and the data are required to be textual, albeit some works designed approaches to integrate structural information such as AST paths or data flow in the input data~\cite{alon2018code2seq, alon2019code2vec, guo2020graphcodebert}. Next, at a higher level, we identified graph-based learning approaches which usually leverage language-specific code abstractions such as ASTs or call graphs. Similarly to natural language processing, here GNNs~\cite{scarselli2008graph, gilmer2017neural} can be used to focus on learning structural information embedded in the input graph data. In addition, graph-based learning approaches benefit from graph representations and their ability to encode higher abstraction and logical structures in a more robust way. Therefore, that makes this latter approach complementary to the two previous one and combining them could result in learning even more potent source code representations by learning from different level of abstractions.

To summarize, we make the following observations: \\
--- \textit{The data that can be extracted from a same software artifact can be highly heterogeneous and lie at different levels of abstraction (\textit{c.f.,} Fig.~\ref{fig:abstraction}), each of them with the potential to offer various structural, semantic and syntactic information about the artifact.} \\
--- \textit{Few amount of previous works propose to combine learning models in a multi-modal fashion by considering several data modalities of software artifacts.}
    
As a matter of fact, to our knowledge multi-modal related work mostly involve bi-modal learning using paired data samples of $<$\textsl{code tokens, code documentation}$>$ in the context of code search or documentation generation~\cite{husain2019codesearchnet, feng2020codebert, guo2020graphcodebert} or multi-modal learning with textual data only~\cite{gu2018deep} and thus without leveraging graph-based modalities that embed more structural information.

These observations lead us to the two following challenges for the future. The first one related to the data and the second one to the learning:
\begin{oframed}
\noindent \underline{Challenge 1}: \textbf{what are meaningful data abstractions to be defined and to extract from source code artifacts in order to enable learning programming knowledge from various data modalities?}

\noindent \underline{Challenge 2}: \textbf{how to efficiently modeling source code artifacts using multi-modal learning approaches?}
\end{oframed}
As part of this work and as we believe the notion of concept to be central in software programming, we make a first attempt to tackle both these challenges by (1) learning representations of concept graphs extracted from code snippets and (2) proposing a learning framework to jointly learn graph representations of the concept graphs and their corresponding code representations. The objective of our approach is to improve the effectiveness of an existing pretrained language model of code in the context of a code search task. In Sec.~\ref{sec:implementation} we provide details about our implementation then present preliminary results in Sec.~\ref{sec:evaluation}. Finally, we discuss future work direction in Sec.~\ref{sec:future} and draw a conclusion.

\section{Code concept graphs-augmented multi-modal learning}
\label{sec:implementation}

\begin{figure*}[!t]
    \centering
    \includegraphics[width=.7\textwidth]{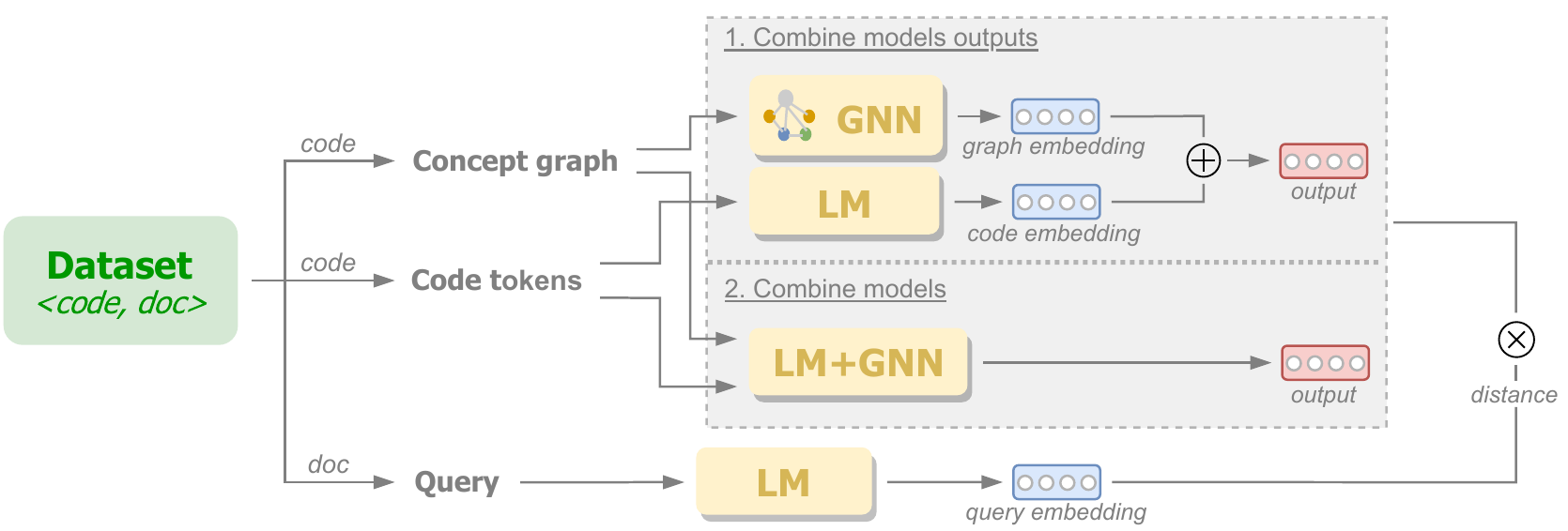}
    \caption{ Overview of our framework. Given a data sample $<$\textsl{code, doc}$>$, we extract a query from the \textsl{doc} and code tokens and a concept graph (CG) from the \textsl{code} (Sec.~\ref{sec:approach_data}). A language model (LM) is used to get a representation of the query and the LM is joint-learned with a concept graph neural network (GNN) with as learning objective to minimize the distance between the representations of the code+CG and the query (Sec.~\ref{sec:approach_learning}). }
    \label{fig:approach}
\end{figure*}

Driven by our claims and motivations, we propose to extract concept graphs from code snippets in view of enhancing an existing pretrained LM of code. By joint-learning a LM based on code tokens and a GNN based on concept graphs, we expect both models to learn better vector representations that incorporate more semantics by covering different level of abstractions thanks to the diversity of the data modalities. Furthermore, as part of our framework, we propose two methods for joint-learning both models. We depict the overall architecture of our framework in Fig.~\ref{fig:approach} and illustrate its applicability to code search.

\subsection{Code concept graphs}
\label{sec:approach_data}

We envision our code concept graphs to represent high-level relationships and dependencies between domain concepts materialized by particular code constructs, \textit{e.g.,} variables, calls, methods. That is, our proposed CG includes only code identifiers and does not directly depend on the syntax of the input language. This is different from other abstractions that have been used previously to improve code comprehension. Among others, some works proposed to create knowledge graphs based on code and documentation~\cite{abdelaziz2020graph4code, cao2019unsupervised} or code ontologies incorporating many linguistic features of the input programming language~\cite{atzeni2017codeontology, jiomekong2019extracting}. On the one hand, the aforementioned works do not intend to integrate their knowledge-based data as part of a deep NLP learning process such as ours. And, conversely to our proposed code CG, the data abstractions in these works are mostly language-specific and do not aim at extracting relational CG of identifiers.

\subsubsection*{Concept graph design}
A code CG is a relational graph $\mathcal{G} = (\mathcal{V}, \mathcal{E})$ where $\mathcal{V}$ denotes the set of identifiers nodes and $\mathcal{E}\subseteq\mathcal{V}\times\mathcal{R}\times\mathcal{V}$ the set of edges with a set of typed relations $\mathcal{R}$. Given a code snippet, we retrieve all the identifiers $(id_{name},\:id_{type})$ where $\textsl{type} \in \{$\textsl{method name, parameters, imports, variables, calls}$\}$ and map it to a unique node $\mathbf{z} \in \mathcal{V}$ with $id_{name}$ and $id_{type}$ its features. Then, we retrieve in total 9 types of relations $\in \mathcal{R}$, each of them describing a semantic dependency  between two identifiers that results from their utilization in the code. We showcase an example for Java in Fig.~\ref{fig:concept_graph} that includes 4 types of relations. For example, ``$dependsOn$'' denotes a relation between the method and an imported identifier that is manipulated in the method. Another example is ``$defines$'' that describes a variable being defined in the body of the method. Or else, ``$calls$'' denotes that a variable depends on the result of a call, \textit{e.g.,} during an assignment. We end-up with a CG of identifiers and their semantic relationships which is also much less dependent on linguistic features of the language than code tokens.

\subsubsection*{Task and data extraction}
Code search is a task that consists of retrieving the most relevant code snippets from a corpus given a natural language query which reflects the user's information need. 
Therefore, the starting point of our approach requires a dataset of pairs of $<$\textsl{code, doc}$>$, \textit{i.e.,} a syntactically valid code method and its corresponding natural language description. Even though the natural language description is not a user query per se, it is commonly used as a proxy query in the literature to train or fine-tune a model without the need of any user relevance's annotations~\cite{gu2018deep, husain2019codesearchnet, feng2020codebert}. From the code snippet, we extract its corresponding CG resulting in triplets of $<$\textsl{CG, code, query}$>$. We choose to explore our approach on Java but our designed CG could be extracted from other programming languages. Additionally, it could be generalized to other tasks with few adjustments.

\subsection{Joint-learning and models combinations}
\label{sec:approach_learning}

Multi-modal or multi-view deep learning refers to learning systems that learn representations from multiple modalities~\cite{ngiam2011multimodal, li2018survey}. In this section, we present two approaches for joint-learning two learning models. We refer to~\cite{li2018survey} for in-depth details on multi-modal representation learning.

\subsubsection*{Combining models outputs}
Lets assume two data modalities whose vector representations say $\mathbf{h}_x$ and $\mathbf{h}_y$ can be obtained by two learning models. The goal of this approach is to obtain a final representation $\mathbf{H}$ = $\phi(\mathbf{h}_x, \mathbf{h}_y)$. The function $\phi$ denotes the combination of both vectors which can be defined as a sum $\phi(\mathbf{h}_x, \mathbf{h}_y) = \mathbf{h}_x + \mathbf{h}_y$, a concatenation $\phi(\mathbf{h}_x, \mathbf{h}_y) = [\mathbf{h}_x, \mathbf{h}_y]$, or a more sophisticated fusion method. Coming back to our data, one can use a LM and GNN with pooling layers to encode respectively the code tokens and the CG into single-vector representations, \textit{i.e.,} code and graph embeddings. The embeddings are then combined using a function $\phi$ to obtain the final representation that embeds knowledge from both modalities.

\subsubsection*{Combining models}
This second more sophisticated method consists of combining modules or sub-parts of the learning models such that their architectures are no longer independent. For instance, existing approaches such as~\cite{kiros2014multimodal} allow to condition the word representations learned by a LM on external modalities such as images. In the context of our work, a LM and a GNN can be combined in several ways. Following~\cite{kiros2014multimodal}, the representation of each identifier code token in the LM could be conditioned on the representation of its corresponding node in the CG obtained by the GNN. Or else, the embedding layer of the LM or the GNN could be initialized with a pretrained embedding layer obtained from the other modality. Conversely to the former one, this method allows more fine-grained combinations of both learning models by integrating the multi-modal learning perspective as part of their architecture. \\

For sake of clarity lets call $\mathbf{h}_q$ the representation of the query, $\mathbf{h}_{code}$ the representation of the code and $\mathbf{h}_{CG}$ the representation of the concept graph. As part of this work, we get the output representation $\mathbf{h}_c$ by combining the models outputs with $\phi = \mathbf{h}_{code} + \mathbf{h}_{CG}$. As the target end-task is code search, the learning objective is to learn representations $\mathbf{h}_q$ and $\mathbf{h}_c$ that are semantically close by minimizing their distance. Therefore, we use cosine similarity to compare $\mathbf{h}_c$ with $\mathbf{h}_q$ and maximize the similarity during learning.

\section{Preliminary evaluation}
\label{sec:evaluation}

To get insights on whether our work direction on multi-modal learning with concept graphs is worth investigating, we conducted a preliminary evaluation of our approach with code search as end-task in Java. 

\subsubsection*{Data and models} 
We use a modified version of CodeSearchNet dataset~\cite{husain2019codesearchnet} including extra preprocessing and cleaning~\cite{guo2020graphcodebert}. It consists of 181,061 pairs of $<$\textsl{code, doc}$>$ with 164,923 for training, 5,183 for validation and 10,955 for testing. From the codes, we extract the data following the process described in Sec.~\ref{sec:approach_data} resulting in triplets of $<$\textsl{CG, code, query}$>$. We use state-of-the-art CodeBERT~\cite{feng2020codebert,codebert-model} as pretrained LM to encode both codes and queries. To encode the CGs, we implemented a gated attention network (GAT)~\cite{gat2018graph, brody2021attentive} with a global attention pooling layer~\cite{li2017gated} on top of it.

\subsubsection*{Models training and test} 
We fine-tune CodeBERT with the $<$\textsl{code tokens, query}$>$ pairs as baseline and joint-learn our CCGNN model, \textit{i.e.,} code concepts GNN, with the pretrained version of CodeBERT, using the triplets $<$\textsl{CG, code, query}$>$. We fine-tune/train both CodeBERT and CCGNN for 10 epochs. To test our models, we use the same evaluation setup as in Husain~\textit{et al.}'s~\cite{husain2019codesearchnet} work where the model has to find the most-relevant code snippet among 1000 candidates given a natural language query.
\begin{figure}[!h]
    \centering
    \includegraphics[width=.78\linewidth]{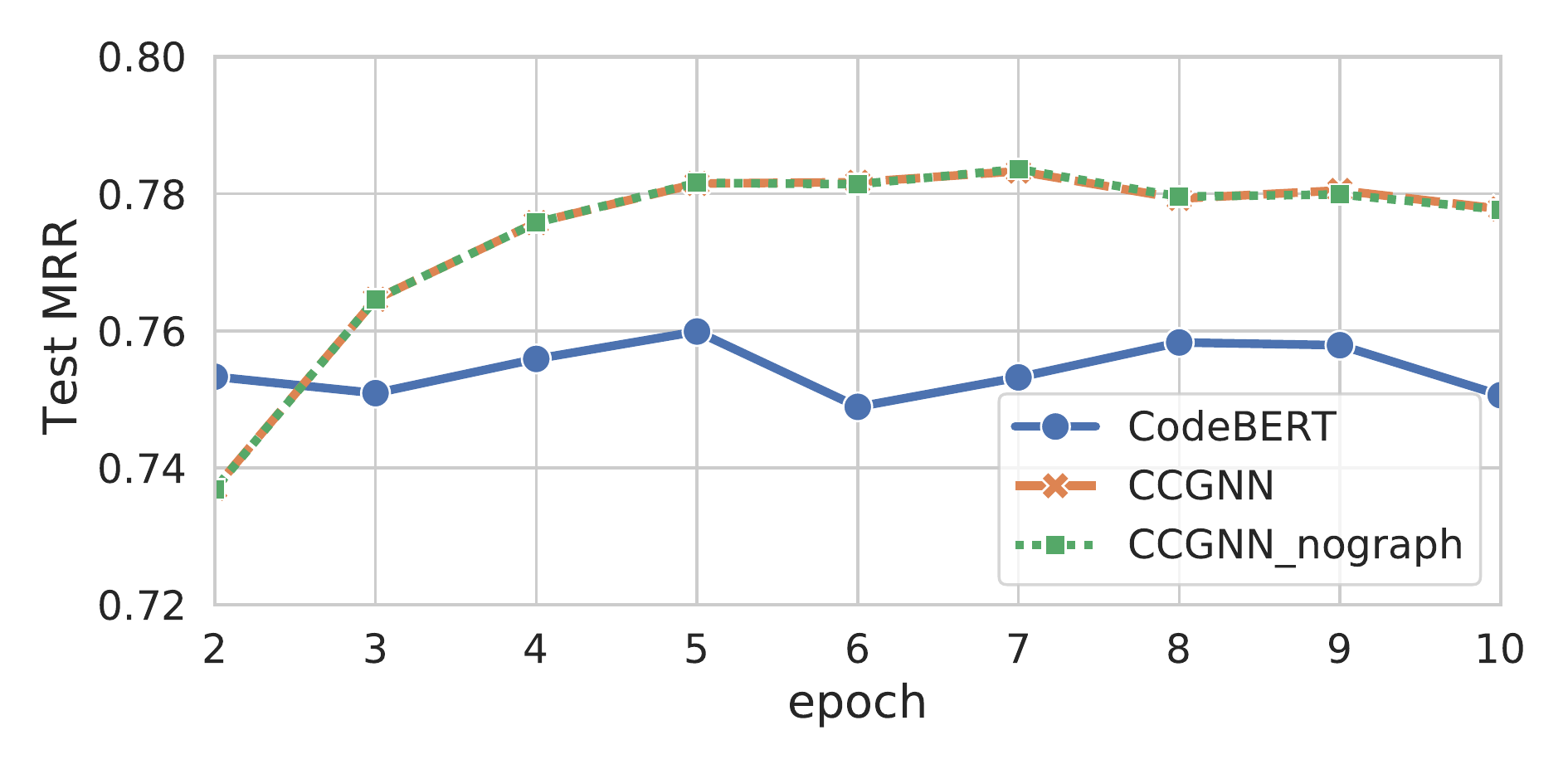}
    \caption{Comparison of the test MRR of CodeBERT and CCGNN (code concept GNN) fine-tuned \textit{w.r.t} the number of training epochs.}
    \label{fig:evaluation}
\end{figure}
\subsubsection*{Results}
In Fig.~\ref{fig:evaluation}, we report the evolution of the mean reciprocal rank (MRR) on the test set for CodeBERT and CCGNN. The third alternative CCGNN\_nograph is equivalent to CCGNN but the CG representations are not used at test time.
We can observe that our approach slightly increases the baseline reaching more than \textbf{0.78} of MRR. However, we performed $t$-tests between CodeBERT and CCGNN for each epoch which showed no significant difference between both models. We did not perform $t$-test between CodeBERT and CCGNN\_nograph as the MRR curve of the latter is identical to CCGNN. This means that in our multi-modal learning approach, the concept graphs representations obtained with the GNN allows to guide the LM of code towards learning better representations. In addition, it does not require to load the CGs and the GNN to achieve better results at test which can be useful for low-computational setups. Finally, even though we show little improvement, we used a very simple combination method that did not require a lot of engineering which might suggest that more sophisticated combination methods could bring impressive gains.

\section{Future plans}
\label{sec:future}

In this preliminary work, we have shown that it is possible to augment an existing powerful pretrained LM of code by incorporating code concept graphs as part of the training process and by joint-learning a LM with a GNN. Besides, our experiments show that our proposal is worth investigating as we achieve better results than a state-of-the-art LM using a simple and straightforward joint-learning method. Consequently, we intend to explore more sophisticated combining methods such as one based on combining the architectures of both learning models instead of combining their outputs. Additionally, even though we illustrate our framework for code search, our code concept graphs-augmented multi-modal learning approach applies to lots of other end-tasks with few adjustments and by modifying the learning objective. Therefore, we envision to design adaptations of our framework to tackle more tasks and evaluate the potential gains that it could bring. And, as our concept graphs intend to be as language agnostic as possible, we envision to extract equivalent CG for more programming languages and leverage transfer learning to apply our framework in several languages. Finally, our proposed framework relies on multi-modal learning and could apply to a lot of other software data modalities. Therefore, we plan on comparing different combination of data modalities in the future.

\section{Conclusion}
\label{sec:conclusion}
In this paper, we present an original idea to abstract knowledge from software using multi-modal learning approaches. The novelty of our vision is twofold. Firstly, it lies on the idea of combining multiple data modalities at different level of abstraction extracted from the same software artifact to modeling the programming world. And secondly, we propose to abstract semantic relationships between identifiers into concept graphs to learn from a high-level structured modality. We conducted a preliminary evaluation of our framework by joint-learning a LM that focuses on learning from code syntax and a GNN leveraging our code concept graphs. The results show gains in the effectiveness of the models for code search using a straightforward combination method of the learning models. Consequently, this preliminary work prompts us to carry on our work in a research line focusing on multi-modal learning approaches for software engineering while keeping a truly data-driven perspective in mind.

\bibliographystyle{ACM-Reference-Format}
\bibliography{references}

\end{document}